\newcommand{\degree}{$^{\circ}$}

\newcommand{\lapp}{$\lesssim$}

\documentclass{pasa}%
\usepackage{graphicx}
\usepackage{amsmath}	
\usepackage{amssymb}	

\title[MWA Tied-Array Processing II]{MWA Tied-Array Processing II: Polarimetric Verification and Analysis of two Bright Southern Pulsars}

\author[Mengyao Xue et al.]{Mengyao Xue$^{1,2}$, S. M. Ord$^{3}$, S. E. Tremblay$^{1,2}$, N. D. R. Bhat$^{1,2}$, C. Sobey$^{1,4}$, B. W. Meyers$^{1,2,3}$, S. J. McSweeney$^{1,2}$, and N. A. Swainston$^{1,2}$
\affil{$^{1}$International Centre for Radio Astronomy Research (ICRAR), Curtin University, Bentley, WA 6102, Australia}%
\affil{$^{2}$ARC Centre of Excellence for All-sky Astrophysics (CAASTRO)}%
\affil{$^{3}$CSIRO Astronomy and Space Science, Australia Telescope National Facility, PO Box 76, Epping, NSW 1710, Australia}%
\affil{$^{4}$CSIRO Astronomy and Space Science, PO Box 1130, Bentley WA 6102, Australia}%
}%

\jid{PASA}
\doi{10.1017/pas.\the\year.xxx}
\jyear{\the\year}

\usepackage{aas_macros}
\usepackage{hyperref}
\usepackage{threeparttablex}
\usepackage{makecell}

\hypersetup{colorlinks,citecolor=blue,linkcolor=blue,urlcolor=blue}

\hypersetup{draft}

\begin{document}

\begin{frontmatter}
\maketitle

\begin{abstract}
Polarimetric studies of pulsars at low radio frequencies provide important observational insights into the pulsar emission mechanism and beam models, and probe the properties of the magneto-ionic interstellar medium (ISM). Aperture arrays are the main form of next-generation low-frequency telescopes, including the Murchison Widefield Array (MWA). These require a distinctly different approach to data processing (e.g. calibration and beamforming) compared to traditional dish antennas. As the second paper of this series, we present a verification of the MWA's pulsar polarimetry capability, using two bright southern pulsars, PSRs J0742--2822 and J1752--2806.
Our observations simultaneously cover multiple frequencies (76--313\,MHz) and were taken at multiple zenith angles during a single night for each pulsar.
We show that the MWA can be reliably calibrated for zenith angles $\lesssim 45$\degree and frequencies $\lesssim 270$\,MHz.
We present the polarimetric profiles for PSRs J0742--2822 and J1752--2806 at frequencies lower than 300\,MHz for the first time, along with an analysis of the linear polarisation degree and pulse profile evolution with frequency.
For PSR J0742--2822, the measured degree of linear polarisation shows a rapid decrease at low frequencies, in contrast with the generally expected trend, which can be attributed to depolarisation effects from small-scale, turbulent, magneto-ionic ISM components. This effect has not been widely explored for pulsars in general, and will be further investigated in future work.

\end{abstract}

\end{frontmatter}

\section{INTRODUCTION}
\label{sec:intro}

Shortly after the original discovery of pulsars \citep{Hewish1968}, the radio emission from pulsars was found to be polarised \citep{Lyne1968}.
The polarimetric profiles of pulsars provide useful tools to classify and study pulsar geometry \citep[e.g. ][]{Rankin1983a,Brinkman2019}.
Most pulsars tend to show a significant amount of linear polarisation -- about 20 per cent, on average, and reaching up to 100 per cent in some cases \citep[e.g.][]{Weltevrede2008,Han2009}.
The degree of linear polarisation generally tends to decrease with increasing observing frequency \citep[e.g.][]{Morris1981,vonHoensbroech1998a}. The physical explanation for this is not straightforward, though may be due to higher frequencies traversing more of the magnetosphere \citep{Johnston2008}, or the birefringence of
plasma in the open field-line region of pulsar magnetospheres \citep[e.g.][]{McKinnon1997}.
Many pulsars also show some circular polarisation -- about 10 per cent, on average \citep[e.g.][]{Gould1998,Weisberg1999,Johnston2018}.

The radio emission from pulsars is broadly thought to originate within the open field lines of the magnetosphere, with the emission beam centred on the magnetic axis \citep[e.g. ][]{Komesaroff1970}.
Consequently, the linear polarisation position angle (PA) will be determined by the direction of the magnetic field line as it sweeps across our line-of-sight. The PA measured as a function of rotational phase, referred to as the PA curve, is generally expected to delineate an S-shape, i.e., the PA will vary more slowly at the outer wings of the pulse profile compared to the centre. This is referred to as the rotating vector model \citep[RVM;][]{Radhak&Cooke1969}. Fitting the RVM model to the observed PAs provides an estimation of the emission beam size and the inclination angle of the magnetic axis with respect to the rotation axis \citep[e.g. ][]{Narayan1982,Everett2001}.

For some pulsars, especially those with medium or low levels of linear polarisation, their PA curves show interruptions of approximately 90\degree, referred to as orthogonal jumps \citep[e.g.][]{McKinnon1998}. For example, \cite{Karastergiou2011} present the polarimetric profiles for PSR J0738--4042 at different epochs, and find that the presence of orthogonal jumps in the PA curves are associated with reduced degrees of linear polarisation. Such observational evidence affirms the concept that pulsar emission could comprise two superposed orthogonal polarisation modes \citep[OPMs; e.g. ][]{McKinnon2000}.
There have been a number of associated studies from both observational \citep[e.g. ][]{Edwards&Stappers2004,Noutsos:2015} and theoretical \citep[e.g. ][]{Gangadhara1997,Melrose2006,vanStraten&Tiburzi2017} points of view. However, the nature of the OPMs is still one of the least understood aspects of pulsar emission.

Of the $>$2600 known pulsars\footnote{ \href{http://www.atnf.csiro.au/people/pulsar/psrcat/}{http://www.atnf.csiro.au/people/pulsar/psrcat/}} \citep{Manchester2005}, only around one-third have polarimetric properties currently available, and this is mostly limited to $\sim$1.4\,GHz frequencies. Extensive polarimetric studies have been published by \cite{Gould1998} using the Lovell telescope, \cite{Weisberg1999} using the Arecibo telescope, \cite{Manchester1998} and \cite{Johnston2018} using the Parkes telescope.
There have been a limited number of such investigations at frequencies below $\sim$1\,GHz. Some studies have been conducted using the Giant Metrewave Radio Telescope {\citep[GMRT;][]{GMRT1991,Roy2010ExA}}, including the Meterwavelength Single-pulse Polarimetric Emission Survey at 333 and 618\,MHz \cite{Mitra2016}. At lower radio frequencies, \cite{Noutsos:2015} undertook polarimetric studies below 200\,MHz using the Low-Frequency Array {\citep[LOFAR;][]{LOFARproject2013,Stappers2011}}, for pulsars observable from the northern hemisphere. For southern pulsars, the Murchison Widefield Array {\citep[MWA;][]{Tingay2013,Wayth2018}} provides an excellent opportunity for polarimetric studies of pulsars at frequencies below $\sim$300\,MHz. Low-frequency polarisation observations of pulsars provide precise probes of the magneto-ionic ISM; for example, towards reconstructing the structure of the Galactic magnetic field \citep{Han2009,Sobey2019}. Furthermore, detailed studies of low-frequency polarimetric profiles, in combination with higher frequency data, can provide further insights into the pulse profile evolution, and potentially help elucidate the magnetospheric radio emission mechanism. They also serve as a useful reference for future pulsar studies planned with the Square Kilometre Array (SKA).

Traditionally, polarimetric studies were mostly performed using large single dish antennas or interferometer telescopes comprised of parabolic dishes. For single dish telescopes like Parkes, polarimetric calibration is achieved by comparing an observed full-Stokes pulse profile to a standard profile template (e.g. using long-term monitoring of PSR J0437--4715). This allows the complex Jones matrix, which describes the transfer function of the signal through the system, to be determined \citep[e.g. ][]{Hamaker1996,Heiles2001,Johnston2002,vanStraten2013}.
For interferometers like the Westerbork Synthesis Radio Telescope (WSRT) or GMRT, the procedure for forming a tied-array beam is to add the voltages of all telescopes after correcting their relative phases. This means that the tied-array signal can be calibrated by determining an overall system Jones matrix, and thus the polarisation calibration process is similar to that of single dishes \citep[e.g. ][]{Edwards&Stappers2004,Mitra2016}.

For low-frequency aperture-array instruments such as LOFAR, the MWA, and the upcoming SKA-Low (i.e. the low frequency component of the SKA), the approach is quite different. Since the arrays have no moving parts, tile/station beam pointings are formed by electronic manipulation of the dipole signals using analogue/digital beamformers \citep{Tingay2013,LOFARproject2013}. This beam response is strongly dependent on the pointing direction and frequency \citep[e.g. ] []{Noutsos:2015} and is more difficult to model compared to that for a single dish.
Accurate calibration of the full polarimetric response is necessary for aperture arrays to facilitate reliable polarisation studies. This procedure is fairly complex, due to a number of factors: the wide field of view; multiple receiving elements; frequency-dependent beam shape; the absence of an injected artificial calibration signal; and the modulation of calibration errors by the time-variable ionosphere. Thus, developing a suitable calibration/beamformation strategy for the MWA and verifying that it is satisfactorily robust and reliable is an essential prerequisite for performing any polarimetric pulsar work with these new generation radio arrays.

In Paper 1 of this series, we described the algorithms and pipeline that we have developed to form the tied-array beam products from the summation of calibrated signals of the antenna elements (Ord et al. submitted).
In this paper, we present the first polarimetric study of two bright southern pulsars, PSRs J0742--2822 and J1752--2806, at frequencies below 300\,MHz. Our primary goal is to investigate their polarimetric properties as a function of frequency and to ascertain the reliability of the MWA's polarimetric performance.
We describe our observations and data processing methods in Section 2. In Section 3 we summarise the optimal calibration strategy and reproducibility of the polarimetric profiles. In Section 4 we describe how we measure the Faraday rotation and use it to estimate the instrumental leakage.
We describe the polarimetric properties of the two pulsars, and compare our profiles with published  polarimetric profiles at higher frequencies, in Section 5. In Section 6, we discuss our results and a summary of our work is given in Section 7.

\section{OBSERVATIONS AND DATA PROCESSING}
\label{sec:data}

Our observations were taken with the Phase I MWA, where the array was comprised of 128 tiles -- each tile consists of 16 dual-polarisation dipole antennas arranged in a regular $4\times 4$ grid, operating at 80--300\,MHz.
The development of the Voltage Capture System \citep[VCS;][]{Tremblay2015} extended the capabilities of the MWA from an imaging interferometer, allowing it to record high-time and -frequency resolution voltage data and enabling time-domain astrophysics. This allows the MWA to provide phase-resolved observations of pulsars \citep[e.g.][]{Bhat2018}.

We use multiple observations of two bright polarised pulsars that pass through the zenith at the MWA site, PSRs J0742--2822 and J1752--2806, to empirically examine the MWA's polarimetric response in the beamforming (tied-array) mode, and its stability across the dipole/antenna beam.
Here, we describe the properties of target pulsars, our observing strategy, calibration, beamformation, data reduction procedures, as well as archival MWA observations used for our comparison studies.

\subsection{Target pulsars}

PSR J0742--2822 has a period ($P$) of 167\,ms, and a dispersion measure (DM; the integrated electron column density along the line-of-sight) of 73.73\,cm$^{-3}$\,pc. It is highly polarised with a linear polarisation degree of around 70\% and a flux density of $\sim$300\,mJy at 400\,MHz \citep{Lorimer1995b,Gould1998,Mitra2016}, and hence a linear polarisation flux density of $\sim$200\,mJy at 400\,MHz.

We note that PSR J0742--2822 is known to exhibit emission mode-changing, however, the timescale is relatively long, typically $\sim$95 days \citep{Keith2013}. All of our observations of PSR J0742--2822 were taken during the same night within five hours (see below) and, therefore, it is likely that all of our data were taken when the pulsar was in the same emission mode. This is consistent with our results (see Section 3.2).

PSR J1752--2806 ($P=563$\,ms, DM=50.37 cm$^{-3}$\,pc) is a moderately polarised pulsar with a linear polarisation degree of around 10\% and a flux density of $\sim$1100\,mJy at 400\,MHz \citep{Lorimer1995b,Gould1998,Mitra2016}; thus, its linear polarisation flux density is $\sim$110\,mJy at 400\,MHz. Typically, flux densities of pulsars follow a power-law with a spectral index of about $-1.6$, and so we expect pulsars to be around three times brighter at 200\,MHz.

\subsection{Observing strategy}
\label{sec:ob_strategy}

For each pulsar, we obtained and compared the polarimetric pulse profiles over a range of zenith angles across a wide range of observing frequencies. These data allowed us to assess the quality, reliability, and repeatability of our calibration and beamformation process.

\begin{figure*}
\begin{center}
\includegraphics[scale=0.12, angle=0]{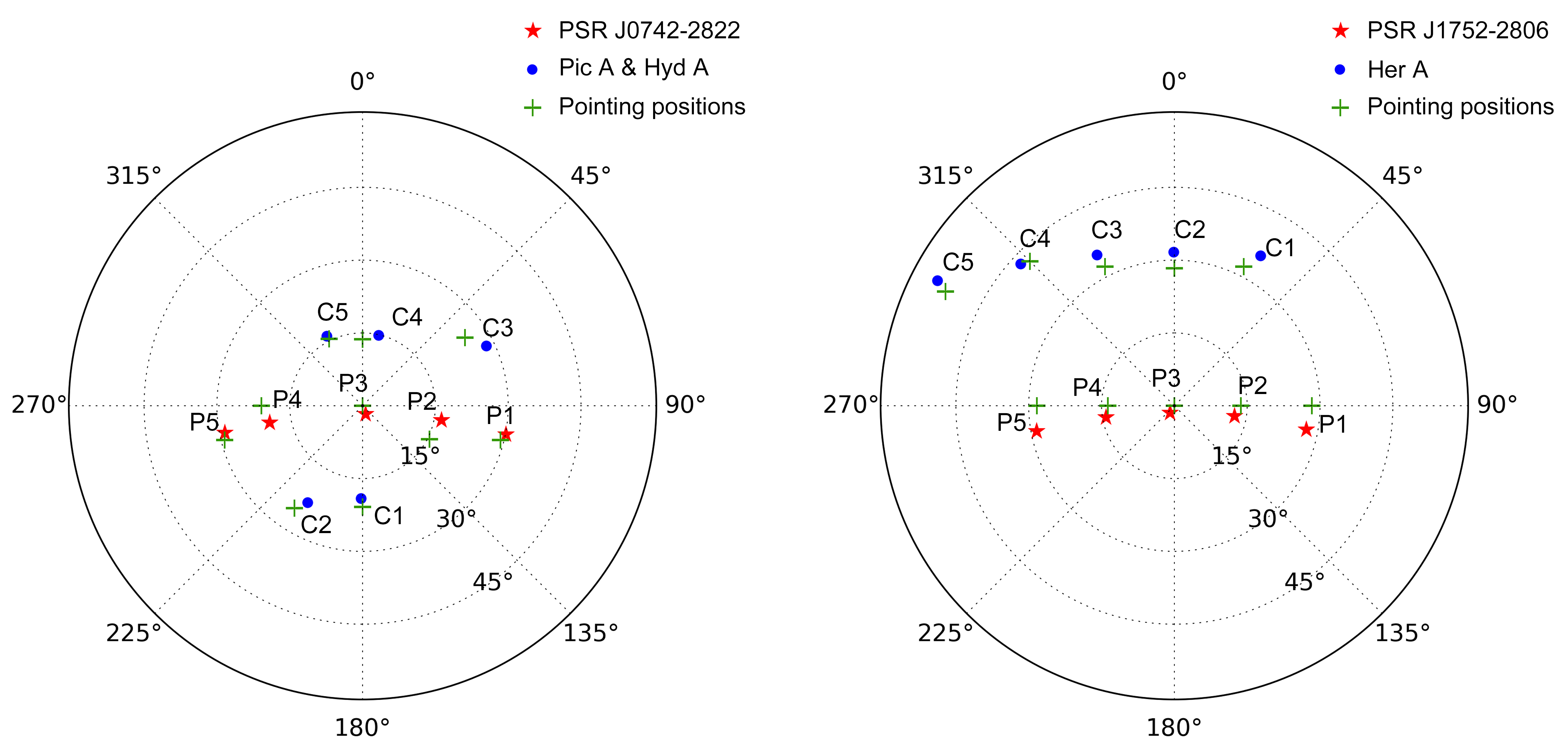}
\caption{(Left) Locations and MWA pointing directions for PSR J0742--2822 and its calibrators Pictor A (C1, C2) and Hydra A (C3--C5). (Right) Locations and MWA pointing directions for PSR J1752--2806 and its calibrator Hercules A. For both panels, the azimuth and zenith angles are shown in horizontal coordinates, azimuth=0\,\degree~represents North and azimuth=90\,\degree~represents East. Red stars indicate the position of pulsars (labelled P1-P5 in order of observation start time); blue circles indicate the position of calibrators (labelled C1-C5 in order of observation start time); green crosses indicate the pointing centre for each observation. Since the MWA points towards `sweet' spots in the sky (dictated by the analogue beamformer settings), there is usually some offset between the pointing centre direction and the position of the target.}\label{Fig:points}
\end{center}
\end{figure*}

Three specific elements of the data were used to investigate the effect on the polarimetric profiles produced, namely:
\begin{itemize}
  \item sky position of the target pulsar;
  \item sky position of the calibrator source;
  \item calibration strategy, i.e., observing a calibrator source prior to the target versus in-field calibration.
\end{itemize}
Consequently, the observing strategy that we adopted is as follows.
For each of the two target pulsars, we recorded five short observations of 5 minutes each, separated by roughly 1 hour, with the middle observation pointing towards the zenith and the others towards zenith angles (ZA) up to 30\degree, as shown in Figure \ref{Fig:points}. The five observations for each pulsar are labelled P1 to P5, in the order of the observing start times.
We also recorded a 5-minute dedicated calibrator observation before each pulsar observation. Thus, there are also five dedicated calibrator observations in total for each pulsar. Similarly, they are named C1 to C5, in the order of their observing start times (see Figure \ref{Fig:points}). The pointing information for the observations is summarised in Table \ref{tab:obs}.

The VCS records data after the second stage of channelisation in the MWA signal path, with a 10\,kHz frequency resolution and a 100\,$\mu$s time resolution.
Data from our observations were recorded at frequencies spread across the entire MWA band, from 76.80\,MHz to 312.32\,MHz, in 24 coarse (1.28\,MHz-wide) channels (see Table \ref{tab:freq}). These non-contiguous coarse channels were recorded simultaneously, each split into 128 fine (10\,kHz-wide) channels.

\begin{table*}[htbp]
  \centering
  \caption{Observing parameters for PSR J0742--2822, PSR J1752--2806, and calibrator sources.} \label{tab:obs}
    \begin{tabular}{l|ccccc}
    \hline\hline
    Target pulsar & \multicolumn{5}{c}{J0742--2822 (observed on 2016-02-04)}\\ \hline
    Observation ID (GPS time) & 1138625088 & 1138628792 & 1138633056 & 1138638456 & 1138641032 \\ 
    Start time (UTC) & 12:44:32 & 13:46:16 & 14:57:20 & 16:27:20 & 17:10:16 \\ 
    (Az, ZA) (deg) & (104.04, 29.26) & (116.56, 15.37) & (0, 0) & (270, 20.83) & (255.96, 29.26) \\ 
    Offset to pointing centre (deg) & 1.6     & 4.6     & 1.8     & 3.7     & 1.5 \\ \hline
    Calibrator & PicA & PicA & HydA & HydA & HydA \\ 
    Observation ID (GPS time) & 1138624768 & 1138628472 & 1138632736 & 1138638136 & 1138640712 \\ 
    Start time (UTC) & 12:39:12 & 13:40:56 & 14:52:00 & 16:22:00 & 17:04:56 \\ 
    (Az, ZA) (deg) & (180, 20.83) & (213.69, 25.31) & (56.31, 25.31) & (0, 13.72) & (333.44, 15.37) \\ 
    Offset to pointing centre (deg) & 1.8     & 2.9     & 4.6     & 3.3     & 0.8 \\
    \hline\hline
    \end{tabular}

    \vspace{24pt}

    \begin{tabular}{l|ccccc}
    \hline\hline
    Target pulsar & \multicolumn{5}{c}{J1752--2806 (observed on 2016-06-10)}\\ \hline
    Observation ID (GPS time) & 1149605152 & 1149609232 & 1149612832 & 1149616432 & 1149620392 \\ 
    Start time (UTC) & 14:45:36 & 15:53:36 & 16:53:36 & 17:53:36 & 18:59:36 \\ 
    (Az, ZA) (deg) & (90, 28.31) & (90, 13.72) & (0, 0) & (270, 13.72) & (270, 28.31) \\ 
    Offset to pointing centre (deg) & 4.8     & 2.5     & 1.7     & 2.3     & 5.0 \\  \hline
    Calibrator & HerA & HerA & HerA & HerA & HerA \\ 
    Observation ID (GPS time) & 1149604832 & 1149608912 & 1149612512 & 1149616112 & 1149620072 \\ 
    Start time (UTC) & 14:40:16 & 15:48:16 & 16:48:16 & 17:48:16 & 18:54:16 \\ 
    (Az, ZA) (deg) & (26.57, 32.02) & (0, 28.31) & (333.44, 32.02) & (315, 42.12) & (296.56, 52.68) \\ 
    Offset to pointing centre (deg) & 4.0     & 3.4     & 2.9     & 1.8    & 2.8 \\
    \hline\hline

    \end{tabular}
\end{table*}

\begin{table}[htbp]
\begin{center}
\caption{Summary of the centre observing frequencies, $f$. The dispersion delay at each frequency for PSRs J0742--2822 and J1752--2806 are presented in milliseconds (${\Delta}t$) and per cent of the pulse period (${\Delta}P$). } \label{tab:freq}
    \begin{tabular}{cccccc}
    \hline\hline
     &  & \multicolumn{2}{c}{J0742$-$2822} & \multicolumn{2}{c}{J1752$-$2806} \\
    channel & $f$ & ${\Delta}t$ & ${\Delta}P$ & ${\Delta}t$ & ${\Delta}P$ \\
     ID & (MHz) & (ms) & (\%) & (ms) & (\%) \\
    \hline
    60    & 76.80 & 13.51 & 8.10  & 9.23  & 1.64 \\
    61    & 78.08 & 12.85 & 7.71  & 8.78  & 1.56 \\
    68    & 87.04 & 9.28  & 5.56  & 6.34  & 1.13 \\
    76    & 97.28 & 6.65  & 3.99  & 4.54  & 0.81 \\
    84    & 107.52 & 4.92  & 2.95  & 3.36  & 0.60 \\
    92    & 117.76 & 3.75  & 2.25  & 2.56  & 0.46 \\
    100   & 128.00 & 2.92  & 1.75  & 1.99  & 0.35 \\
    116   & 148.48 & 1.87  & 1.12  & 1.28  & 0.23 \\
    117   & 149.76 & 1.82  & 1.09  & 1.24  & 0.22 \\
    124   & 158.72 & 1.53  & 0.92  & 1.05  & 0.19 \\
    132   & 168.96 & 1.27  & 0.76  & 0.87  & 0.15 \\
    140   & 179.20 & 1.06  & 0.64  & 0.73  & 0.13 \\
    148   & 189.44 & 0.90  & 0.54  & 0.61  & 0.11 \\
    156   & 199.68 & 0.77  & 0.46  & 0.53  & 0.09 \\
    164   & 209.92 & 0.66  & 0.40  & 0.45  & 0.08 \\
    165   & 211.20 & 0.65  & 0.39  & 0.44  & 0.08 \\
    172   & 220.16 & 0.57  & 0.34  & 0.39  & 0.07 \\
    180   & 230.40 & 0.50  & 0.30  & 0.34  & 0.06 \\
    212   & 271.36 & 0.31  & 0.18  & 0.21  & 0.04 \\
    220   & 281.60 & 0.27  & 0.16  & 0.19  & 0.03 \\
    228   & 291.84 & 0.25  & 0.15  & 0.17  & 0.03 \\
    236   & 302.08 & 0.22  & 0.13  & 0.15  & 0.03 \\
    243   & 311.04 & 0.20  & 0.12  & 0.14  & 0.02 \\
    244   & 312.32 & 0.20  & 0.12  & 0.14  & 0.02 \\

    \hline\hline
    \end{tabular} \\
\end{center}
\end{table}

\subsection{Calibration}

In order to coherently sum the signals from the MWA tiles, we need to calibrate the array to determine the direction-independent complex gains (amplitudes and phases) for each tile first. The Real Time System \citep[RTS;][]{Mitchell2008} was used for this calibration process. The RTS generated solutions for each tile at each frequency band from calibrator observation. This process utilised the MWA's analytical beam model.
A more detailed description of the calibration process can be found in Section 2.4 of Paper 1.

Each dipole antenna in the MWA array has a beam response, which effectively `point' towards the zenith, and each tile (comprising 4$\times$4 dipoles) has analogue beamformers that add delays to the signals to steer the `tile beam' towards the target source. We can coherently sum the signals from all 128 tiles to form a tied-array beam. In this sense, all of our calibrations can be considered as 'off-axis'.

For the verification described in this paper, we used both the dedicated calibrator observations (as described in Section \ref{sec:ob_strategy}), as well as in-field calibration. For in-field (self) calibration, we processed the complex voltage data from the pulsar observations using an offline version of the MWA correlator \citep{Ord2015} to generate visibilities in the same format as the dedicated calibrator observations.

\subsection{Beamforming}

A detailed description of the beamforming process is presented in Paper 1. Here, we briefly summarise the process.
We used the calibration solutions to coherently combine the signals from all 128 tiles to form a tied-array (phased array) beam towards the target pulsar by applying phase rotations to correct for geometric delays between the tiles so that the entire array `points' towards the target position. This procedure involves applying direction-independent phase and amplitude corrections from the calibration solutions to the signals received by each tile. Meanwhile, since the calibration solution is direction independent, we rotate the calibration solution towards the target position according to the analytic tile beam model (the same as that used in the RTS calibration process), to retrieve the direction-dependent complex gain components before we apply it in the beamforming operation.

\subsection{Data reduction and analysis}
\label{sec:data-reduction}

The beamforming pipeline writes out the data in the PSRFITS format \citep{Hotan2004}. These data were then incoherently de-dispersed and folded using \texttt{DSPSR}\footnote{http://dspsr.sourceforge.net/} \citep{vanStraten2011dspsr} and the timing ephemeris from the pulsar catalogue$^{1}$ \citep{Manchester2005}. For further analysis we mostly used the \texttt{PSRCHIVE}\footnote{http://psrchive.sourceforge.net/} software \citep{Hotan2004,vanStraten2012psrchive}. The optimal DM (maximising the signal-to-noise ratio in the total intensity pulse profile) was found using the PSRCHIVE \texttt{pdmp} routine. The total degree of linear polarisation is estimated as equivalent integrated on-pulse flux of the linearly polarised profile\footnote{The residual off-pulse baseline is estimated and subtracted in our analysis.}. This is essentially the continuum equivalent quantities as it is estimated from the integrated on-pulse profile.

To obtain the polarimetric profiles for the pulsars and estimate the degree of linear polarisation, we need to correct for the effect of Faraday rotation.
We determined the Faraday rotation measure (RM, in units of rad\,m$^{-2}$) towards the pulsars using the technique of RM Synthesis \citep{Burn1966,Brentjens2005}.
We used the pre-processing parts of the \texttt{rmfit} quadratic fitting function within the \texttt{PSRCHIVE} package \citep{Noutsos2008} to extract the Stokes $I, Q, U, V$ parameters (as a function of frequency) from the peak of the average total intensity pulse profile. The Stokes parameters were used as an input to the RM Synthesis code written in \texttt{python}\footnote{https://github.com/gheald/RMtoolkit}.
We also used the associated \texttt{RM-CLEAN}$^{4}$ method \citep{Heald2009,Michilli2018} to determine the fraction of instrumental polarisation leakage near 0\,rad\,m$^{-2}$.
We obtained the dirty and RM CLEAN-ed Faraday spectra (or Faraday dispersion functions) output for each pulsar observation. The RM was recorded as the location of the peak in the clean Faraday spectrum.

The ionosphere imparts additional Faraday rotation, RM$_{\rm{ion}}$, to the total observed RM, RM$_{\rm{obs}}$. The ionospheric contribution must be subtracted in order to obtain the RM due to the interstellar medium (ISM) alone, i.e., RM$_{\rm{ISM}}$=RM$_{\rm{obs}}$--RM$_{\rm{ion}}$. The RM$_{\rm{ion}}$ is both time and direction dependent and, was estimated towards each target source line-of-sight (at the corresponding ionospheric pierce point) for the average time of each observation. To estimate RM$_{\rm{ion}}$, we employed an updated version of \texttt{ionFR}\footnote{https://sourceforge.net/projects/ionfarrot/} \citep{Sotomayor-Beltran2013} using input data from: the latest version of the International Geomagnetic Reference Field\footnote{https://www.ngdc.noaa.gov/IAGA/vmod/igrf.html} \citep[IGRF-12;][]{Thebault2015}; and the International Global Navigation Satellite Systems Service (IGS) vertical total electron content (TEC) maps\footnote{ftp://cddis.gsfc.nasa.gov/pub/gps/products/ionex/} \citep[e.g.][]{Hernandez-Pajares2009} for each observation date.
Using the Long Wavelength Array, \cite{Malins2018} find that local high-cadence TEC measurements are superior to the global TEC models for ionospheric RM correction. However, the difference between the local measurements and the global models is not significant for our range of observing frequencies.

\subsection{Archival VCS data}

In addition to multiple short observations using non-contiguous frequency channels, we also make use of archival VCS data for PSR J1752--2805 for own comparison studies. Specifically, a 40-minute zenith-pointing observation from June 2015 (observation ID: 1117643248), over a contiguous 30.72\,MHz bandwidth at a central frequency of 118.40\,MHz. We used an observation of a calibrator source, 3C444, taken six hours later that night (observation ID: 1117643248, with azimuth=288.43 deg, ZA=22.02 deg) to calibrate and beamform the data using the same procedure as described above.

\section{Verification of the polarimetric response}
\label{sec:results}

Theoretically, the polarimetric response including leakage for the array can be estimated using the MWA beam models and the intrinsic cross-polarization ratio (IXR) \citep{Carozzi2011}. In practice, it is difficult to estimate this accurately due to the discrepancy between the actual beam response and our beam models. Although significant efforts have been made to improve our understanding of the MWA's actual beam response \citep[e.g.][]{Sutinjo2015,Sokolowski2017,Line2018}, it is still an ongoing area of research. Any discrepancy between the actual beam performance and the assumed beam model will manifest as errors in the polarimetric response that are a function of pointing direction. As described in Paper 1, the current MWA tied-array data processing (calibration and beamforming) still uses a simple analytical beam model.
Here, we estimate the polarisation leakage empirically by comparing the results from each target pulsar at different sky coordinates (Az, ZA) calibrated with solutions from multiple (Az, ZA).
Through this comparison, we first determine the optimal calibration strategy (Section 3.1), and then estimate the reproducibility of the polarimetric profiles for the two target pulsars (Section 3.2). Given the large range of instrumental polarisation exhibited by a phased array as a function of pointing, any small errors in the calibration process will manifest as large errors in measured polarisation. We therefore use the reproducibility of the pulsar polarimetric profile over a wide range of instrumental polarisation as a proxy for calibratibility.

\subsection{Determining the optimal calibration strategy}
In order to determine the optimal calibration strategy, different calibration solutions were used to beamform multiple pointings for both pulsars.
For PSR J0742$-$2822, the in-field calibration and five dedicated calibrations are compared for each of the five short pulsar observations. However, for PSR J1752$-$2806, we were unable to benefit from in-field calibration because its line-of-sight is closer to that of the Galactic centre ($l=1.54^{\circ}$, $b=-0.96^{\circ}$), where the bright, diffuse sky background makes it difficult for the calibration procedure to converge on satisfactory solutions. At frequencies above 270\,MHz, where the MWA beam model is not well determined, in-field calibration may not necessarily converge to satisfactory solution for PSR J0742$-$2822, which is located away from the Galactic centre ($l=243.77^{\circ}$, $b=-2.44^{\circ}$).

\begin{figure*}
\begin{center}
\includegraphics[scale=0.16, angle=0]{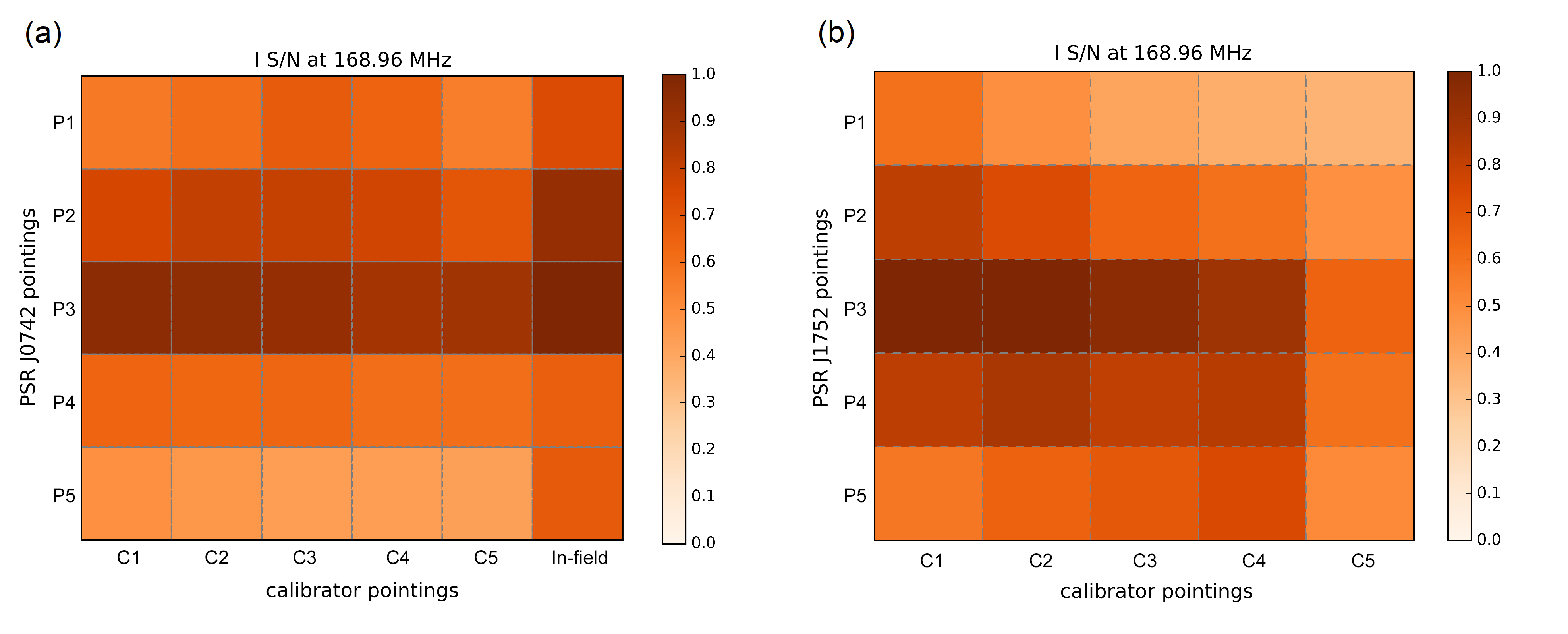}
\caption{Signal-to-noise ratios of the total intensity pulse profiles at 168.96\,MHz with a bandwidth of 1.28\,MHz, for each of the five pulsar pointings calibrated using each of the five dedicated calibrator observations. All values are normalised to the maximum signal-to-noise ratio among the 25 combinations. (a) PSR J0742--2822 observations, (b) PSR J1752--2806 observations. Note that in-field calibration was not performed for PSR J1752--2806 (see Section 3.1 for details).}\label{Fig:pcolor1}
\end{center}
\end{figure*}

\begin{figure*}
\begin{center}
\includegraphics[scale=0.16, angle=0]{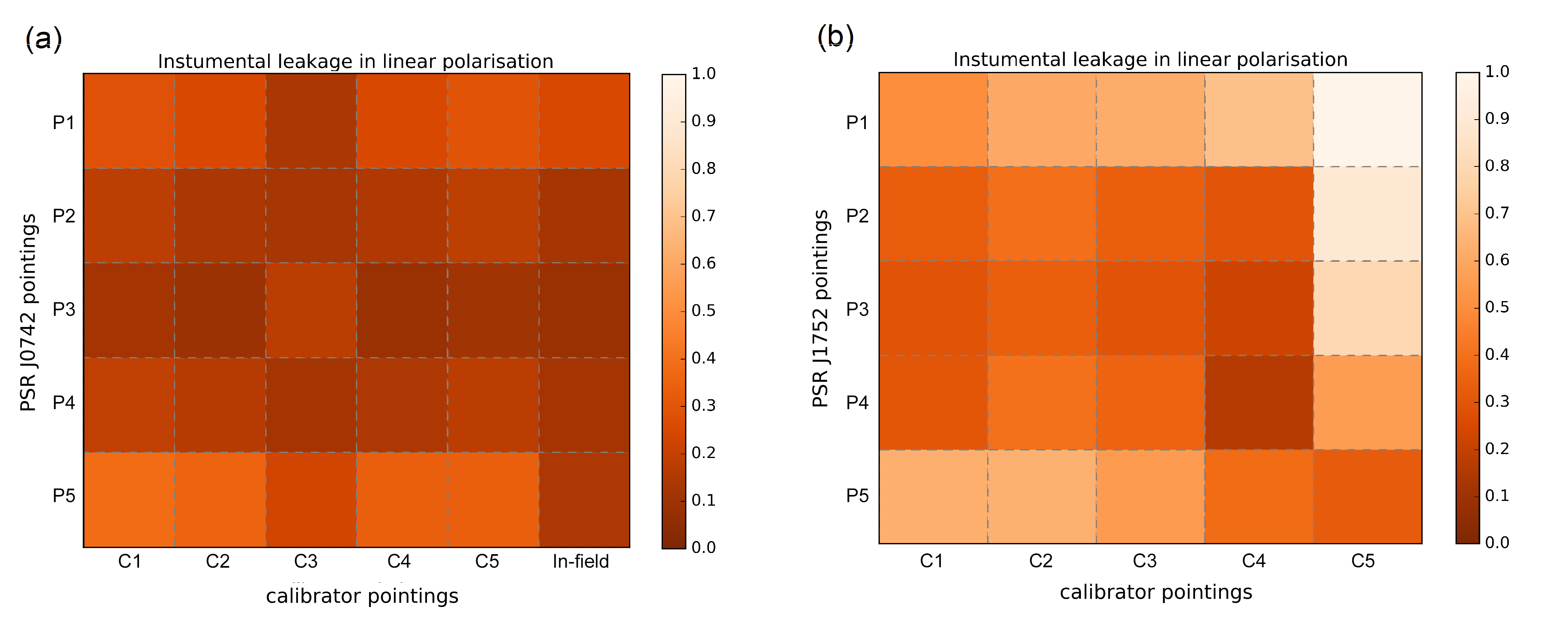}
\caption{Fraction of instrumental polarisation leakage calculated using the RM CLEAN-ed Faraday spectra, for each of the five pulsar pointings calibrated using each of the five dedicated calibrator observations. (a) PSR J0742--2822 observations, (b) PSR J1752--2806 observations.} \label{Fig:pcolor2}
\end{center}
\end{figure*}

The visualisations shown in Figures \ref{Fig:pcolor1} and \ref{Fig:pcolor2} provide comparisons of the signal-to-noise ratio in total intensity (Figure \ref{Fig:pcolor1}) and the instrumental polarisation leakage (Figure \ref{Fig:pcolor2}) for each pulsar. We present all combinations of the pulsar pointings (P1 to P5) and calibration observations (C1 to C5, and in-field calibration for PSR J0742--2822) for one example observing frequency. More details regarding the Faraday spectra can be found in sections \ref{sec:data-reduction} and \ref{sec:RM}.
For PSR J0742$-$2822, we found that the in-field calibration is most effective, yielding higher signal-to-noise ratios in total intensity (Figure \ref{Fig:pcolor1}a) and slightly lower instrumental polarisation fractions (Figure \ref{Fig:pcolor2}a). This is consistent with the conclusion from the MWA image processing \citep{Trott2016,Hurley-Walker2017}.
For PSR J1752$-$2806, the calibrator observations (C1 to C4) all have satisfactory total intensity signal-to-noise ratios, except for the fifth calibrator observation (C5) that is located at a zenith angle of $>$50\,\degree. The fourth calibrator observation shows somewhat less instrumental leakage (Figure \ref{Fig:pcolor2}b), and this may due to the symmetrical projection for both X and Y dipoles at the pointing location.

In summary, the in-field calibration performs well when both the sky model and the MWA beam model are sufficiently well determined. Towards lines-of-sight where this is not necessarily true, the dedicated calibrator observation is sufficient. We note that these are preliminary results based on two case studies, and a more detailed analysis involving a larger sample of pulsars would be useful to develop a higher degree of confidence. However, such an exercise is beyond the scope of this paper.

\subsection{Reproducibility of pulse profiles}

Here, we describe our analysis of the MWA tied-array data at different directions (Az, ZA), which shows that the polarimetric response is satisfactorily reliable for science, such as that demonstrated in the subsequent sections.
The total intensities of the pulse profiles are found to be relatively stable, with a $\sim$10\% variation in signal-to-noise, especially for calibrator observations where the source was less than 45\degree from the zenith, e.g., Figures \ref{Fig:pcolor1} and \ref{Fig:pcolor2}.
The zenith pointing provides the pulse profile with the highest signal-to-noise ratio. This is consistent with our expectation because the MWA has maximum gain and the beam model is more accurately determined for pointings toward the zenith (except at frequencies higher than 270\,MHz, see Section \ref{sec:Discussleakage}). The degree of linear polarisation (over the five different pointing positions and for each observing frequency between 150--270\,MHz) varies by 8$\pm$5\% for PSR J0742--2822 and 4$\pm$2\% for PSR J1752--2806. Some of this discrepancy can be attributed to intrinsic pulse-to-pulse variations that contribute to the average pulse profile. The number of pulses within the short 5-minute observations is $\sim$1800 for PSR J0742--2822 and $\sim$500 for PSR J1752--2806. More than approximately 1000 pulses are typically required to construct average pulse profiles that are stable, depending on the pulse phase-jitter properties of the pulsar \citep[e.g.][]{Liu2012}. Nevertheless, the profiles are also qualitatively similar, with the difference between the profiles attributable to the signal-to-noise ratios.

\section{Faraday rotation measures}
\label{sec:RM}

We calculated the RM towards PSR J0742$-$2822 for all five observations separately (using the \texttt{RM-CLEAN} routine) to investigate the change in the apparent RM at different observing epochs and pointing directions. We also calculated the ionospheric Faraday rotation contribution for each observation. The results are summarised in Table\,\ref{tab:RM1}. Before subtracting the ionospheric RM, the observed RM for PSR J0742$-$2822 ranged from 149.065 to 149.512\,rad\,m$^{-2}$ with a formal error of $\sim$0.05\,rad\,m$^{-2}$. After subtracting the ionospheric RM, the average RM$_\text{ISM}$ for PSR J0742$-$2822 was estimated to be 150.915$\pm$0.097\,rad\,m$^{-2}$. Thus, at these low frequencies, it is essential to correct for the ionospheric Faraday rotation, but the measurement uncertainty is dominated by the method currently used.
Similarly, we determined the RM towards PSR J1752$-$2806 for all five five-minute observations and also for the 40-minute observation centred at 118.40\,MHz. Before subtracting the ionospheric RM, the observed RM ranged from 95.001 to 95.068\,rad\,m$^{-2}$. The measurement uncertainties for the five short observations are around 0.015\,rad\,m$^{-2}$, while for the observation centred at 118.40\,MHz, the measurement uncertainty is only 0.01\,rad\,m$^{-2}$. The results are also summarised in Table\,\ref{tab:RM1}.

\begin{figure*}
\begin{center}
\includegraphics[scale=0.28]{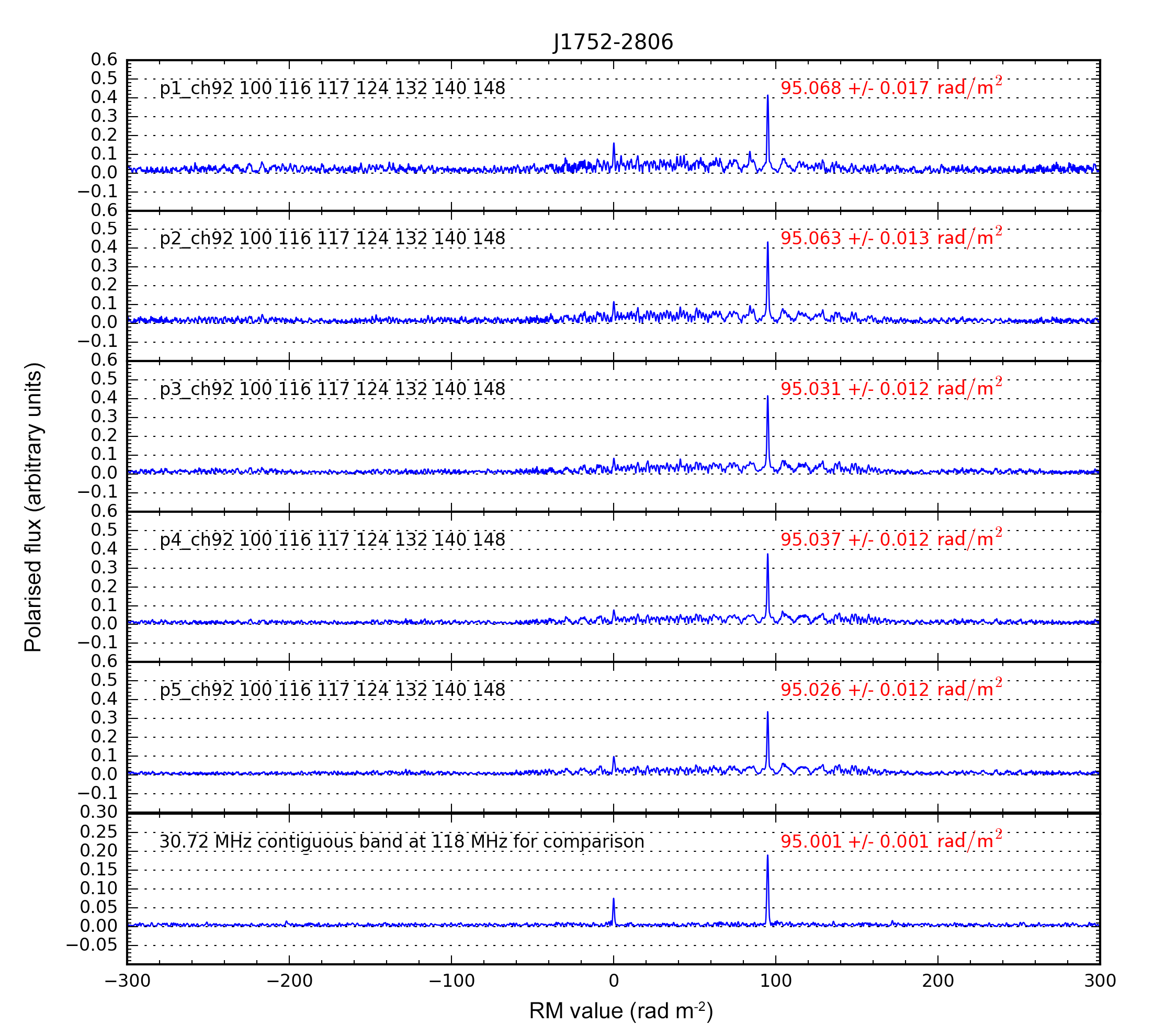}
\caption{RM CLEAN-ed Faraday spectra for PSR J1752$-$2806. The labels in red show the observed RM prior to the subtraction of the ionospheric RM, and the uncertainty quoted is the formal error. The upper five panels show the Faraday spectra obtained from multiple short observations with observing frequencies 117.76--189.44\,MHz. The bottom panel is the result from the observation centred at 118.40\,MHz.} \label{Fig:RMclean1}
\end{center}
\end{figure*}

\begin{figure*}
\begin{center}
\includegraphics[scale=0.28]{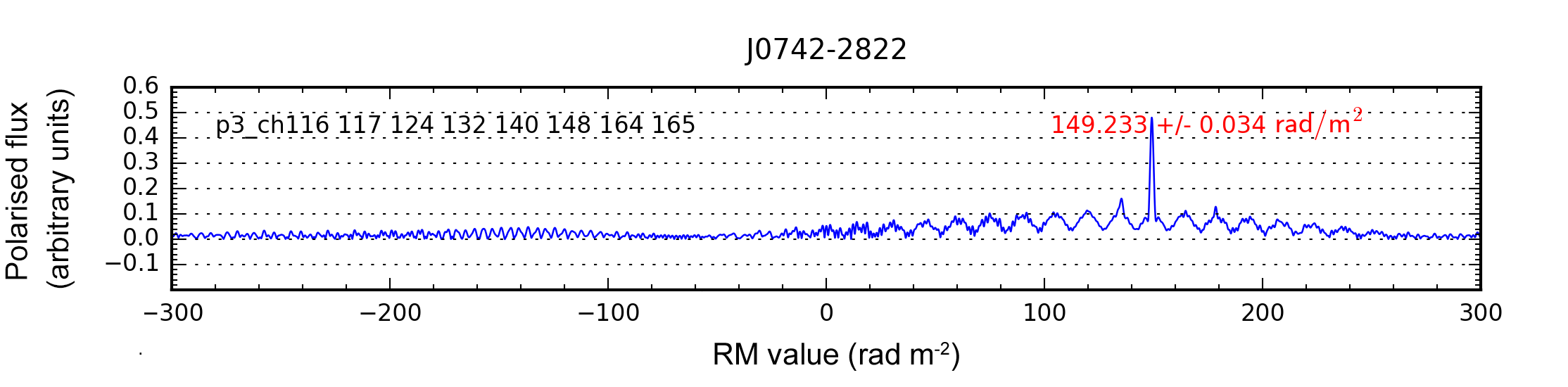}
\caption{RM CLEAN-ed Faraday spectrum for PSR J0742$-$2822 using the third short observation (P3) with observing frequencies 148.48--211.20\,MHz. The label in red is the observed RM prior to the subtraction of the ionospheric RM, and the uncertainty quoted is the formal error.} \label{Fig:RMclean2}
\end{center}
\end{figure*}

Figure \ref{Fig:RMclean1} shows the RM CLEAN-ed Faraday spectra for PSR J1752$-$2806 from all five short observations and the 118.40\,MHz observation. The height of the peak at 0\,rad\,m$^{-2}$, relative to the higher peak at the RM associated with the pulsar signal, indicates the fraction of instrumental polarisation leakage. We also show an example of an RM CLEAN-ed Faraday spectrum for the third short observation of PSR J0742$-$2822 in Figure \ref{Fig:RMclean2}.

We note that the RM measurement uncertainty is proportional to the full width at half maximum (FWHM) of the rotation measure spread function (RMSF; analogous to a PSF in optical telescopes) and inversely proportional to the signal-to-noise ratio in the Faraday spectrum.
The use of low-frequencies and wide bandwidths reduces the FWHM of the RMSF, thus reducing the uncertainty on the RM measurement. At our lowest observing frequencies the pulsar profiles have lower signal-to-noise ratios, while at higher frequencies (particularly above 200\,MHz), the instrumental polarisation leakage becomes greater (see Section \ref{sec:Discussleakage}).
Including more of these higher-frequency channels increases the systematic error because the peaks associated with the pulsar signal and the instrumental polarisation in the Faraday spectrum calculated using each coarse frequency channel overlap due to the wider FWHM of the RMSF and increasing instrumental polarisation leakage. To achieve a balance between attaining smaller RM uncertainties by maximising the frequency range used, while reducing systematic errors from including high-frequency channels and noisier low-frequency channels, we adopted the frequency range 148.48--211.20\,MHz to calculate the RM for PSR J0742$-$2822 and 117.76--189.44\,MHz for PSR J1752$-$2806 (see Figures \ref{Fig:m1} and \ref{Fig:m2}). For each pulsar, we calculated the Faraday spectrum for each coarse frequency channel individually to assist in deciding these optimum frequency ranges. We note that the RM measurements obtained for each channel (listed in Table \ref{tab:freq}) were all consistent within the formal errors.

Using RM synthesis and the MWA's large frequency lever arm, we are able to achieve higher precision on the RMs compared to previous studies at higher frequencies published in the literature. For example, the RM measured towards PSR J1752--2806 using the MWA data is more than twice as precise and in excellent agreement (0.5-$\sigma$) with the value reported in the pulsar catalogue from \cite{Hamilton1987}. As is evident from Table 3, the largest contribution to the RM uncertainty is the accuracy with which the ionospheric RM can be determined and corrected.
Therefore, ongoing efforts to increase the accuracy of the ionospheric RM estimation are essential to fully realise the potential of high precision of RM measurements from low-frequency instruments \citep[e.g.][]{Malins2018}.

\begin{table*}[htbp]\small
  \centering
  \begin{threeparttable}
  \caption{RM results for PSRs J0742$-$2822 and J1752$-$2806.} \label{tab:RM1}
    \begin{tabular}{l|ccccc}
    \hline\hline
    Target pulsar & \multicolumn{5}{c}{J0742--2822 (Literature RM=149.95$\pm$0.05\tnote{\textdagger} ~from \citealt{Johnston2005})}\\ \hline
    Observation ID (GPS time) & 1138625088 & 1138628792 & 1138633056 & 1138638456 & 1138641032 \\ 
    Observed RM (rad\,m$^{-2}$) & 149.065$\pm$0.055 & 149.215$\pm$0.042 & 149.233$\pm$0.034 & 149.512$\pm$0.045 & 149.443$\pm$0.058 \\ 
    Ionosphere RM (rad\,m$^{-2}$) & $-$1.975$\pm$0.144 & $-$1.672$\pm$0.132 & $-$1.466$\pm$0.055 & $-$1.509$\pm$0.042 & $-$1.485$\pm$0.034 \\ 
    RM$_\text{ISM}$ = RM$_\text{obs}$ - RM$_\text{ion}$ & 151.040$\pm$0.154 & 150.887$\pm$0.139 & 150.699$\pm$0.065 & 151.021$\pm$0.062 & 150.928$\pm$0.067 \\ \hline
    Average RM$_\text{ISM}$ (rad\,m$^{-2}$) & \multicolumn{5}{c}{150.915$\pm$0.097} \\
    \hline\hline
    \end{tabular}

    \vspace{24pt}

    \begin{tabular}{l|cccccc}
    \hline\hline
    Target pulsar & \multicolumn{6}{c}{J1752--2806 (Literature RM=96.0$\pm$0.2 from \citealt{Hamilton1987})}\\ \hline
    Observation ID (GPS time) & 1149605152 & 1149609232 & 1149612832 & 1149616432 & 1149620392 & 1117643248 \\ 
    Observed RM (rad\,m$^{-2}$)& 95.068$\pm$0.017 & 95.063$\pm$0.013 & 95.031$\pm$0.012 & 95.037$\pm$0.012 & 95.026$\pm$0.012 & 95.001$\pm$0.001  \\ 
    Ionosphere RM (rad\,m$^{-2}$)& $-$0.886$\pm$0.082 & $-$0.828$\pm$0.072 & $-$0.789$\pm$0.038 & $-$0.877$\pm$0.080 & $-$0.815$\pm$0.083 & $-$0.807$\pm$0.104 \\ 
    RM$_\text{ISM}$ = RM$_\text{obs}$ - RM$_\text{ion}$ & 95.954$\pm$0.084 & 95.891$\pm$0.073 & 95.820$\pm$0.040 & 95.914$\pm$0.081 & 95.841$\pm$0.084 & 95.808$\pm$0.104 \\ \hline
    Average RM$_\text{ISM}$ (rad\,m$^{-2}$) & \multicolumn{5}{c}{95.871$\pm$0.078} \\
    \hline\hline
    \end{tabular}

    \begin{tablenotes}
        \item[\textdagger] This value is the observed RM, with no correction for the ionospheric RM value.
    \end{tablenotes}
  \end{threeparttable}
\end{table*}

\section{Polarimetric profiles}

\subsection{PSR J0742$-$2822}

\begin{figure*}
\begin{center}
\includegraphics[scale=0.8, angle=0]{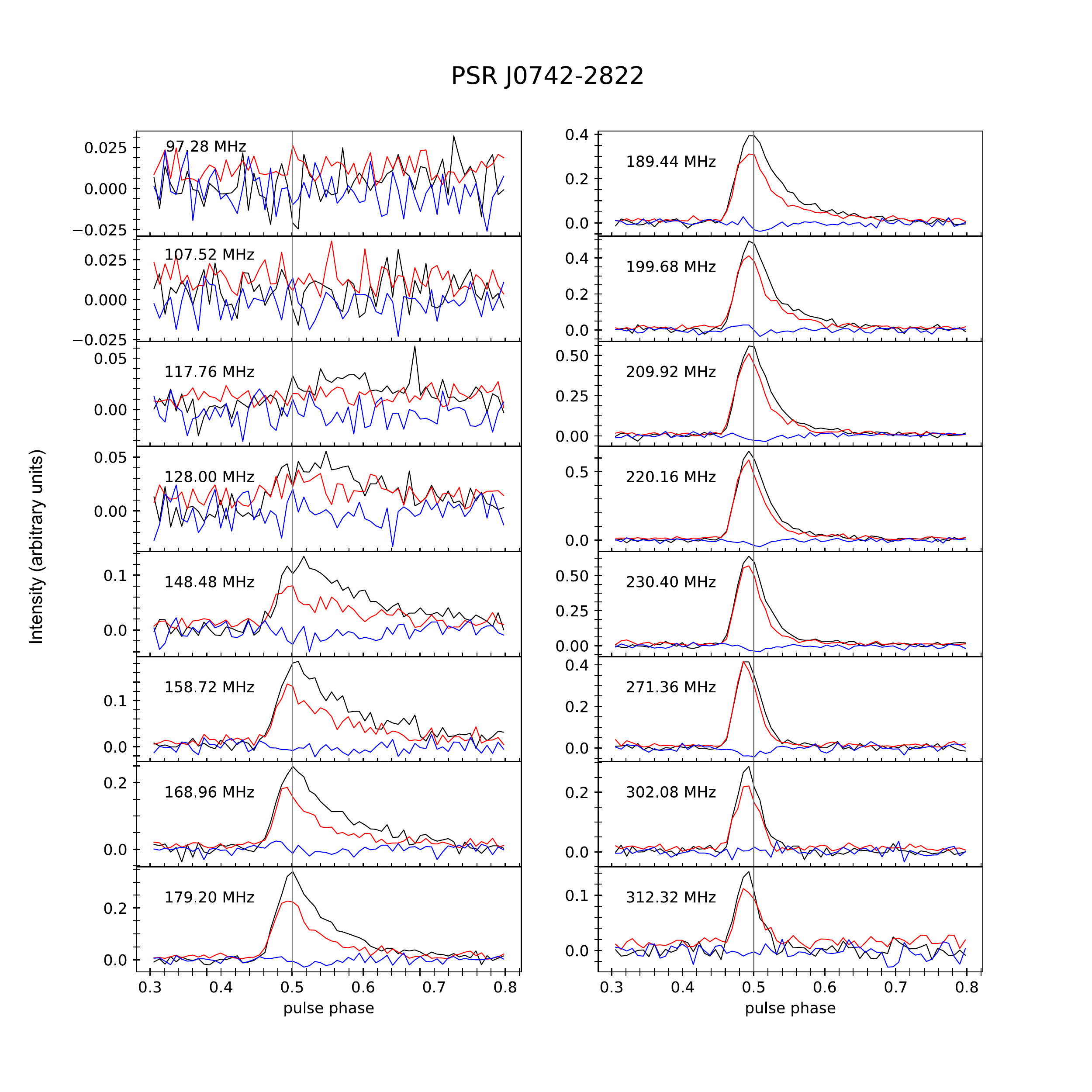}
\caption{Polarimetric profiles for PSR J0742$-$2822 at frequencies from 97\,MHz to 313\,MHz, each with a bandwidth of 1.28\,MHz. Black lines indicate total intensity (Stokes $I$), red lines indicate linear polarisation (Stokes $\sqrt{Q^{2}+U^{2}}$), and blue lines indicate circular polarisation (Stokes $V$). The data used here are the combination of all five short observations of PSR J0742$-$2822 and calibrated using the fifth calibration observation (C5). Note that the polarisation profiles above 270\,MHz are increasingly affected by the instrumental polarisation leakage.}\label{Fig:m1}
\end{center}
\end{figure*}

We obtained polarimetric profiles for our target pulsars for a number of frequencies within the MWA's operational frequency range.
For PSR J0742$-$2822, the pulse profiles at the majority of the observing frequencies show significant exponential scattering tails, see Figure \ref{Fig:m1}.
The reduced telescope sensitivity and the increasing effects of scattering makes it difficult to detect the pulsar at frequencies below 128\,MHz. A detailed analysis of scattering properties of this pulsar can be found in \cite{Kirsten2019}.
Despite the notable scattering, the pulse profile is highly linearly polarised at all frequencies, although the linear polarisation degree decreases towards lower frequencies.
This may suggest that the scattering in the ISM gives rise to depolarisation of the pulsar emission; see Section \ref{sec:depol} for further discussion.

\begin{figure}
\begin{center}
\includegraphics[scale=0.4, angle=0]{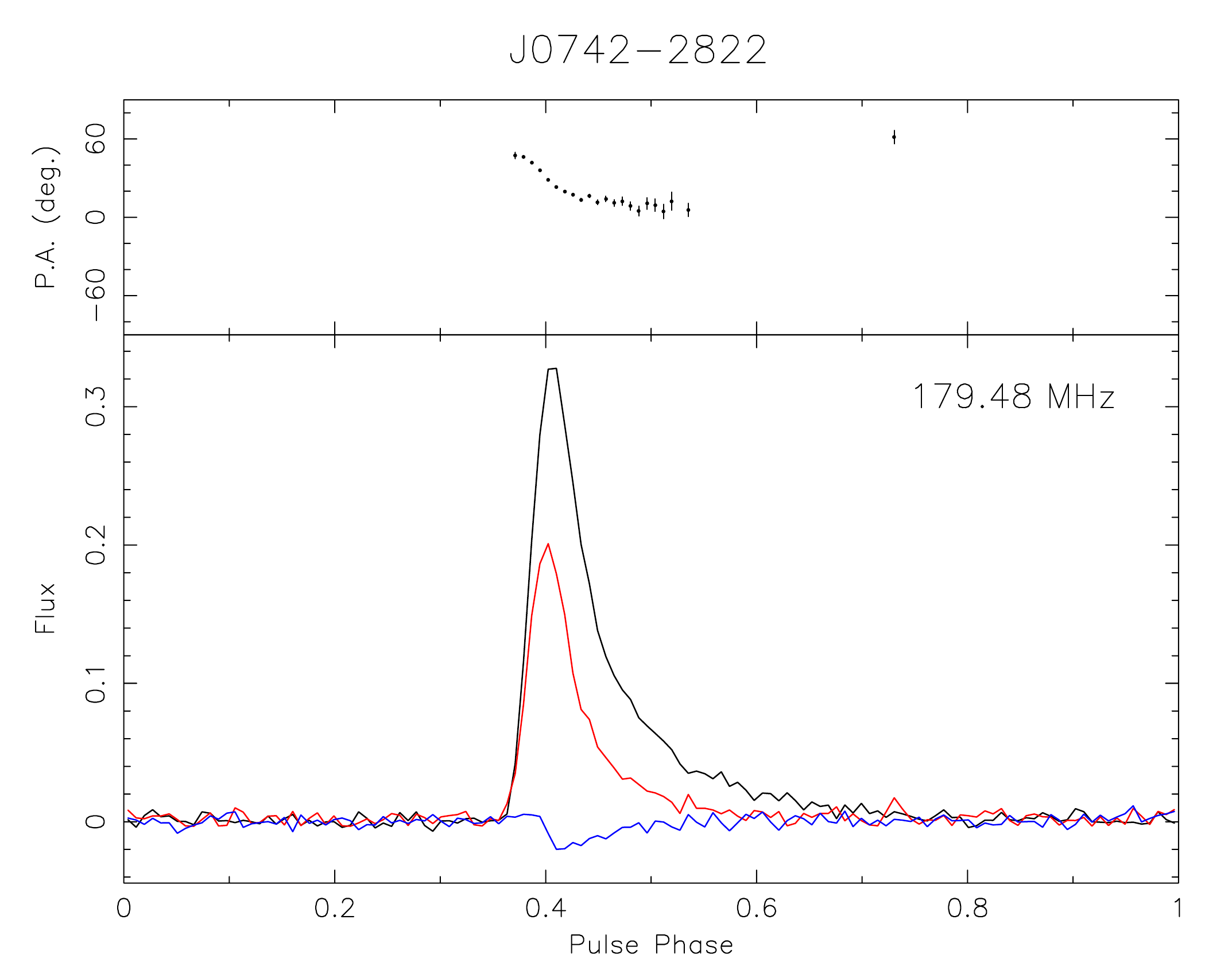}
\caption{Polarimetric profile (lower panel) and PA curve (upper panel) for PSR J0742$-$2822. The black, red, and blue lines indicate the total intensity, linear polarisation, and circular polarisation, respectively. Here, we use the addition of nine 1.28\,MHz channels between 148\,MHz and 211\,MHz. The profile is generated using the data from all five observations, and calibrated using the fifth calibrator observation. The flux density is shown in arbitrary units.}\label{Fig:PA1}
\end{center}
\end{figure}

Figure \ref{Fig:PA1} shows the polarimetric profile of PSR J0742--2822 generated using the combined data from all five short observations over the frequency range 148\,MHz to 212\,MHz (i.e. 9$\times$1.28\,MHz coarse channels). The PA curve shows a $\sim$50\degree swing and flattens towards later pulse phases in the scattering tail.

PSR J0742--2822 was also blindly detected in an all-sky survey of circular polarisation at 200\,MHz with the MWA by \cite{Lenc2018}.
Their measured degree of circular polarisation, $\approx -10$\%, is comparable with our estimation, $-7\pm18$\%. This result is also consistent with the $\approx -7$\% circular polarisation from the Lovell telescope observation at 230\,MHz \citep{Gould1998}.

\subsection{PSR J1752$-$2806}

\begin{figure*}
\begin{center}
\includegraphics[scale=0.8, angle=0]{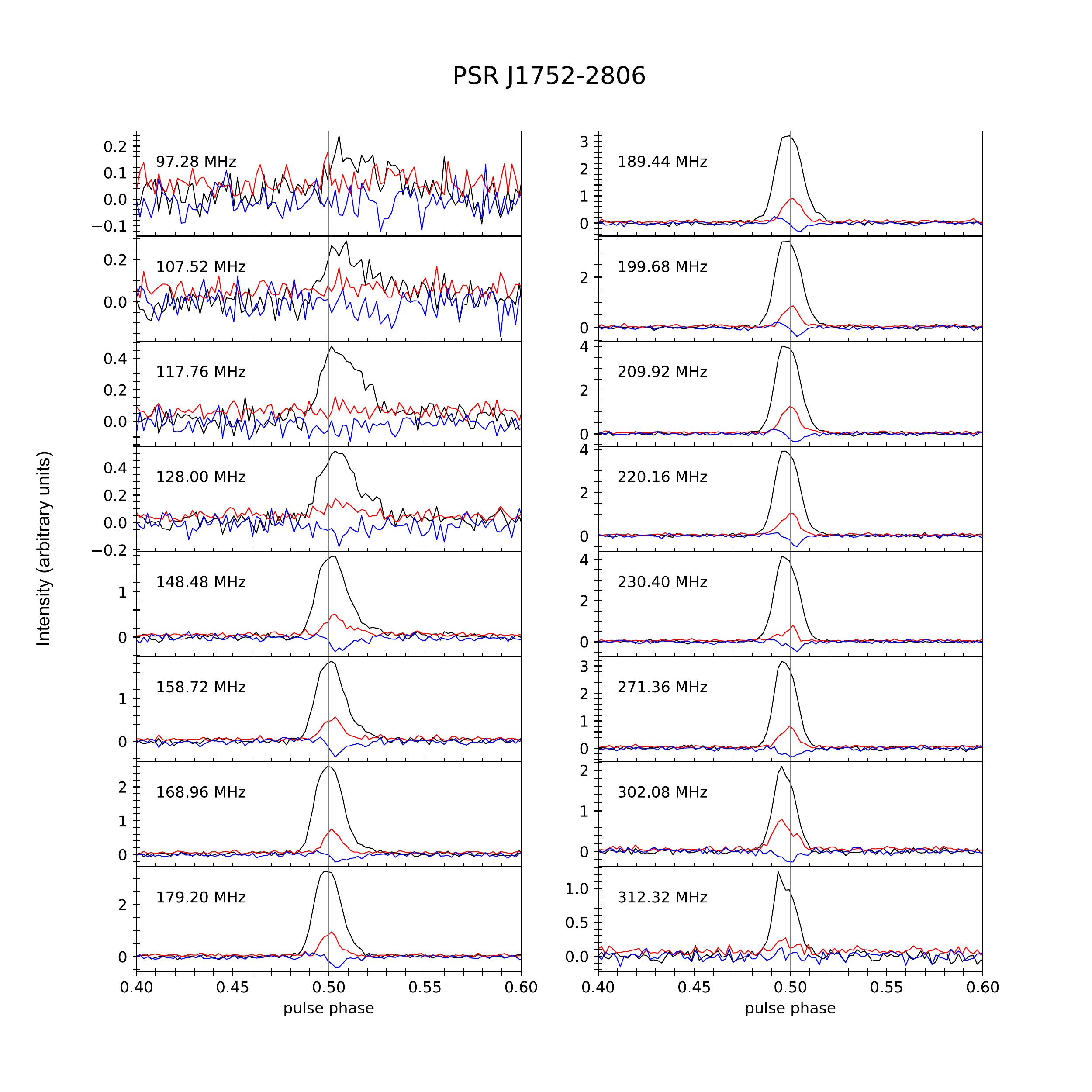}
\caption{Polarimetric profiles for PSR J1752$-$2806 at frequencies from 97\,MHz to 313\,MHz, each with a bandwidth of 1.28\,MHz. Black lines indicate total intensity (Stokes $I$), red lines indicate linear polarisation (Stokes $\sqrt{Q^{2}+U^{2}}$), and blue lines indicate circular polarisation (Stokes $V$). The data used here is the combination of all five short observations of PSR J1752$-$2806 and calibrated using the fourth calibration observation (C4). Note that the polarisation profiles above 270\,MHz are increasingly affected by the instrumental polarisation leakage.}\label{Fig:m2}
\end{center}
\end{figure*}

Figure \ref{Fig:m2} shows the pulse profiles for PSR J1752$-$2806 from 97\,MHz to 312\,MHz. The polarimetric profiles of PSR J1752--2806 show little evolution between 148\,MHz and 220\,MHz, even though measurably longer scattering tails are seen towards the lower frequencies. The signal-to-noise ratio is comparatively lower at frequencies below 118\,MHz. The profiles at 230\,MHz and 312\,MHz show reduced linear polarisation while those at 271\,MHz and 302\,MHz become dominated by instrumental leakage. This may be due to the increased instrumental leakage at frequencies above $\approx$270\,MHz \citep[cf.][]{Sutinjo2015}, see Section \ref{sec:Discussleakage} for further discussion.

\begin{figure*}
\begin{center}
\includegraphics[scale=0.4, angle=0]{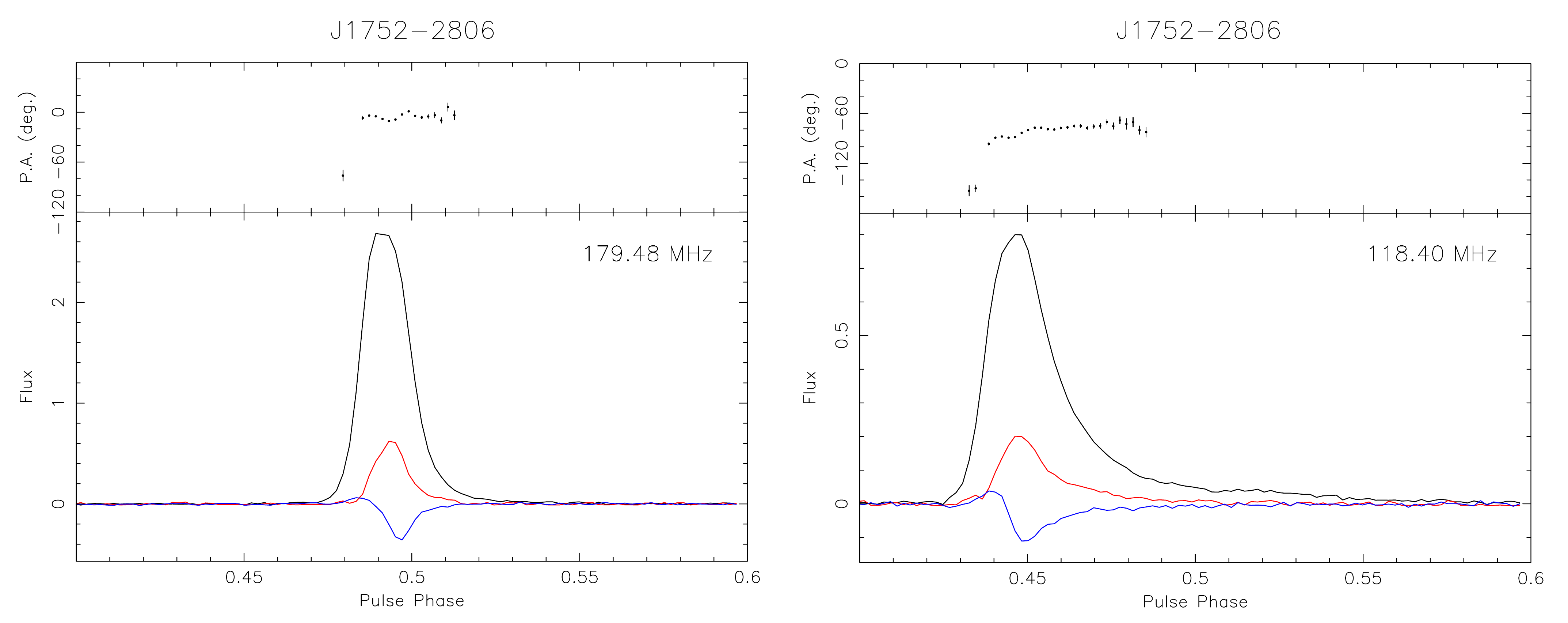}
\caption{Polarimetric pulse profiles (lower panel) and PA curves (upper panel) for PSR J1752$-$2806. The black, red, and blue lines indicate the total intensity, linear polarisation, and circular polarisation, respectively. The flux densities are in arbitrary units. Left: the average profile for all five short observations for PSR J1752$-$2806, calibrated using the fourth calibrator observation (C4). Here, we use the addition of nine 1.28\,MHz channels between 148\,MHz and 211\,MHz. Right: the average profile from the MWA data centred at 118\,MHz with 30.72\,MHz contiguous bandwidth.}\label{Fig:PA2}
\end{center}
\end{figure*}

Figure \ref{Fig:PA2} shows the polarimetric profiles for PSR J1752$-$2806 using the MWA data centred at 180\,MHz and 118\,MHz. There appears to be a jump in the PA curve near the leading edge of the pulse profile, at both centre frequencies. This jump is $\sim 68 \pm 8$ deg at 180\,MHz and $\sim 59 \pm 5$ deg at 118\,MHz.
This jump also tentatively appears to coincide with a faint linear polarisation component at the leading edge of the pulse profile which is rather difficult to discern, given the time resolution of our data (see Table 2). Additional observations high quality low-frequency observations will be useful to further investigate this.

A closer examination of the polarimetric profiles at multiple MWA frequencies (Figure \ref{Fig:m2}) reveals that the scattering effect is negligible for this pulsar above $\sim$150\,MHz. However, as can be seen from Figure \ref{Fig:PA2} (right panel), the pulse broadening arising from scattering is visibly strong at 118\,MHz. Using the deconvolution method \citep[as described in ][]{Bhat2003}, we estimate the scatter broadening time to be 7.7$\pm$0.7 ms at 118\,MHz (assuming a one-side exponential for the pulse broadening function).

\subsection{Comparison with the literature}
\label{sec:Literature}
\begin{figure*}
\begin{center}
\includegraphics[scale=0.25, angle=0]{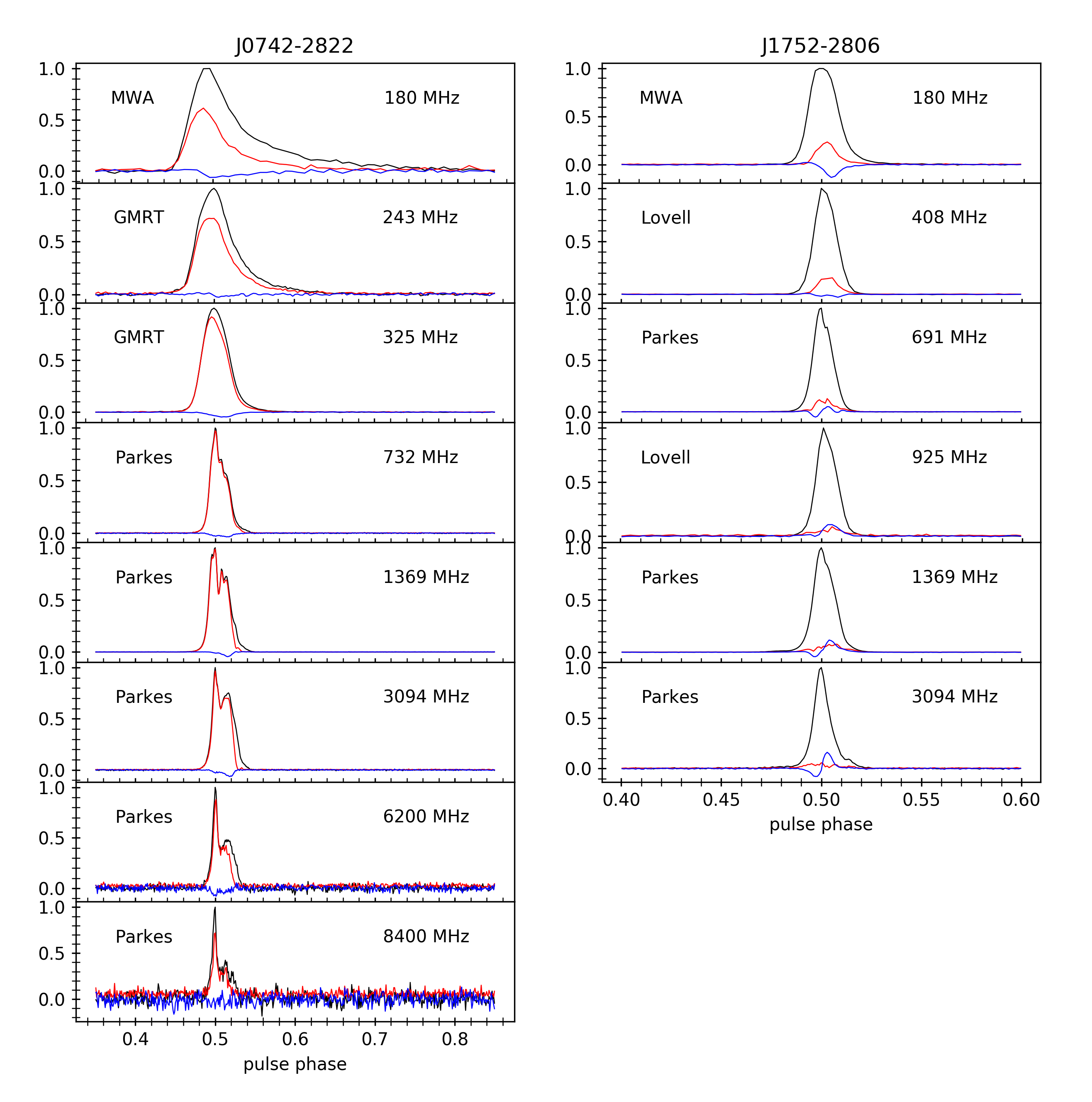}
\caption{Polarimetric profiles for PSR J0742$-$2822 (left) and J1752$-$2806 (right) from the MWA and the literature \citep{Gould1998,Karastergiou2006,Johnston2006,Johnston2007,Johnston2008}. Black lines indicate the total intensity (Stokes $I$), red lines indicate linear polarisation (Stokes $\sqrt{Q^{2}+U^{2}}$), and blue lines indicate circular polarisation (Stokes $V$). The flux density of each of the profiles is normalised (arbitrary units). The labels indicate the telescopes used to observe the pulsars and the centre observing frequencies.}\label{Fig:m3}
\end{center}
\end{figure*}

Figure \ref{Fig:m3} shows the comparison between the polarimetric pulse profiles we obtained using the MWA and those available in the published literature.
The MWA provides the lowest frequency pulse profiles. The profiles published in the literature provide additional data at frequencies up to 8400\,MHz for PSR J0742--2822 and up to 3100\,MHz for PSR J1752--2806.

For PSR J0742$-$2822, we obtained polarimetric profiles taken with the GMRT at 243\,MHz and 325\,MHz \citep{Johnston2008} and Parkes profiles at 732\,MHz, 1.4\,GHz, 3.1\,GHz, 6.2\,GHz, and 8.4\,GHz \citep{Karastergiou2006,Johnston2006}.
All of these are published in the literature, except for the 243, 325, 732, and 6200\,MHz data\footnote{www.atnf.csiro.au/people/Simon.Johnston/ppdata/}.
Pulse broadening (scattering) arising from multi-path propagation in the ISM can be clearly seen in the MWA and GMRT profiles.
The emission is also highly linearly polarised, with a notable decrease towards lower frequencies; see Section \ref{sec:depol} for further discussion.

For PSR J1752$-$2806, the literature archival data include those from the Lovell telescope \citep[408\,MHz and 925\,MHz; ][]{Gould1998} and Parkes \citep[691\,MHz, 1.4\,GHz, 3.1\,GHz; ][]{Karastergiou2006,Johnston2007}.
The linear polarisation fraction is smaller ($<$20\%), compared to PSR J0742--2822, showing some profile evolution with increasing degrees of polarisation towards lower frequencies. In addition, the circular polarisation fraction is often comparable to that for the linear polarisation, and also shows some evolution with frequency.

\section{DISCUSSION}
\label{sec:discussion}

\subsection{Frequency-dependent degree of linear polarisation}
\label{sec:depol}

\begin{figure*}
\begin{center}
\includegraphics[scale=0.5, angle=0]{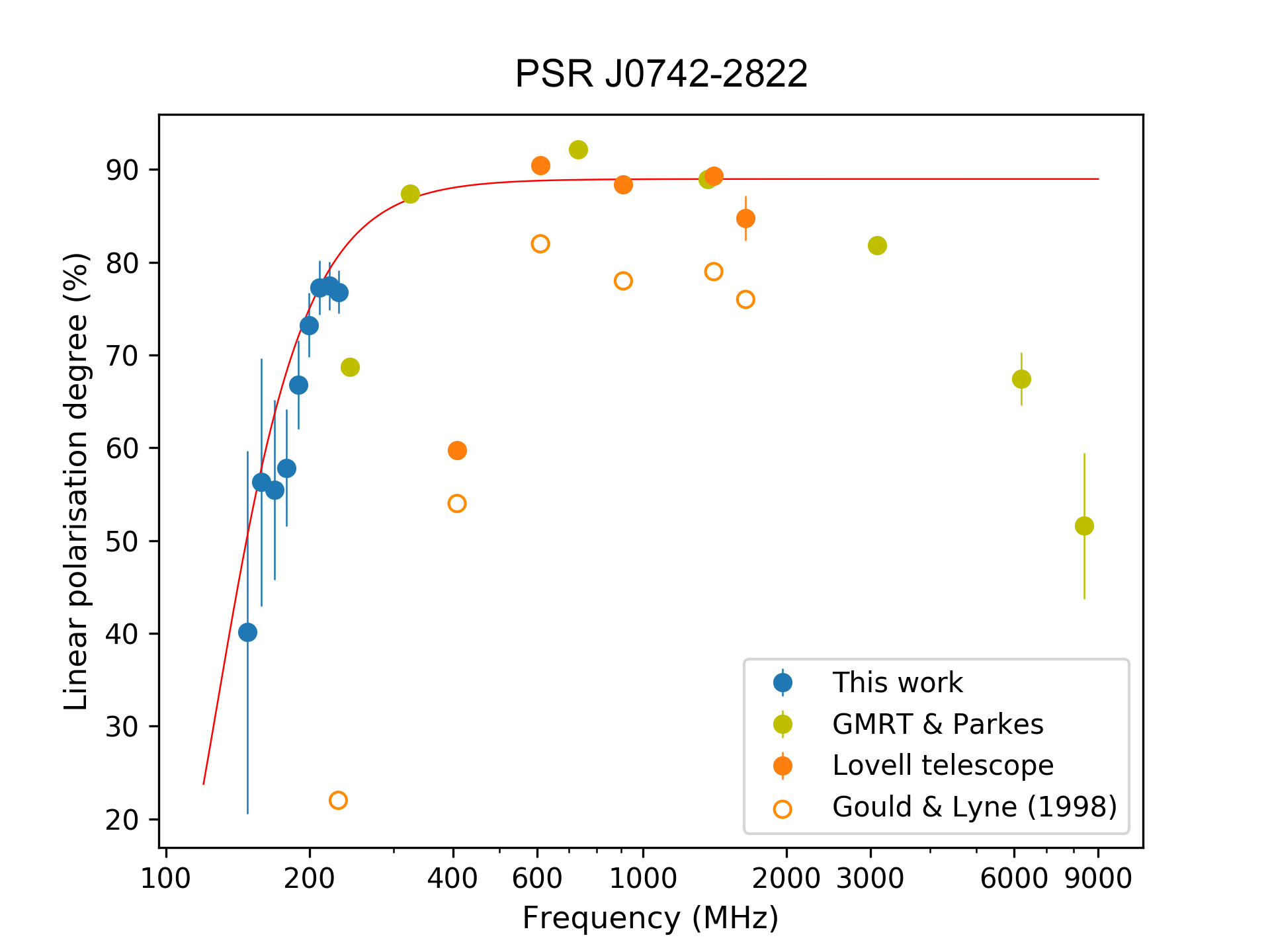}
\includegraphics[scale=0.5, angle=0]{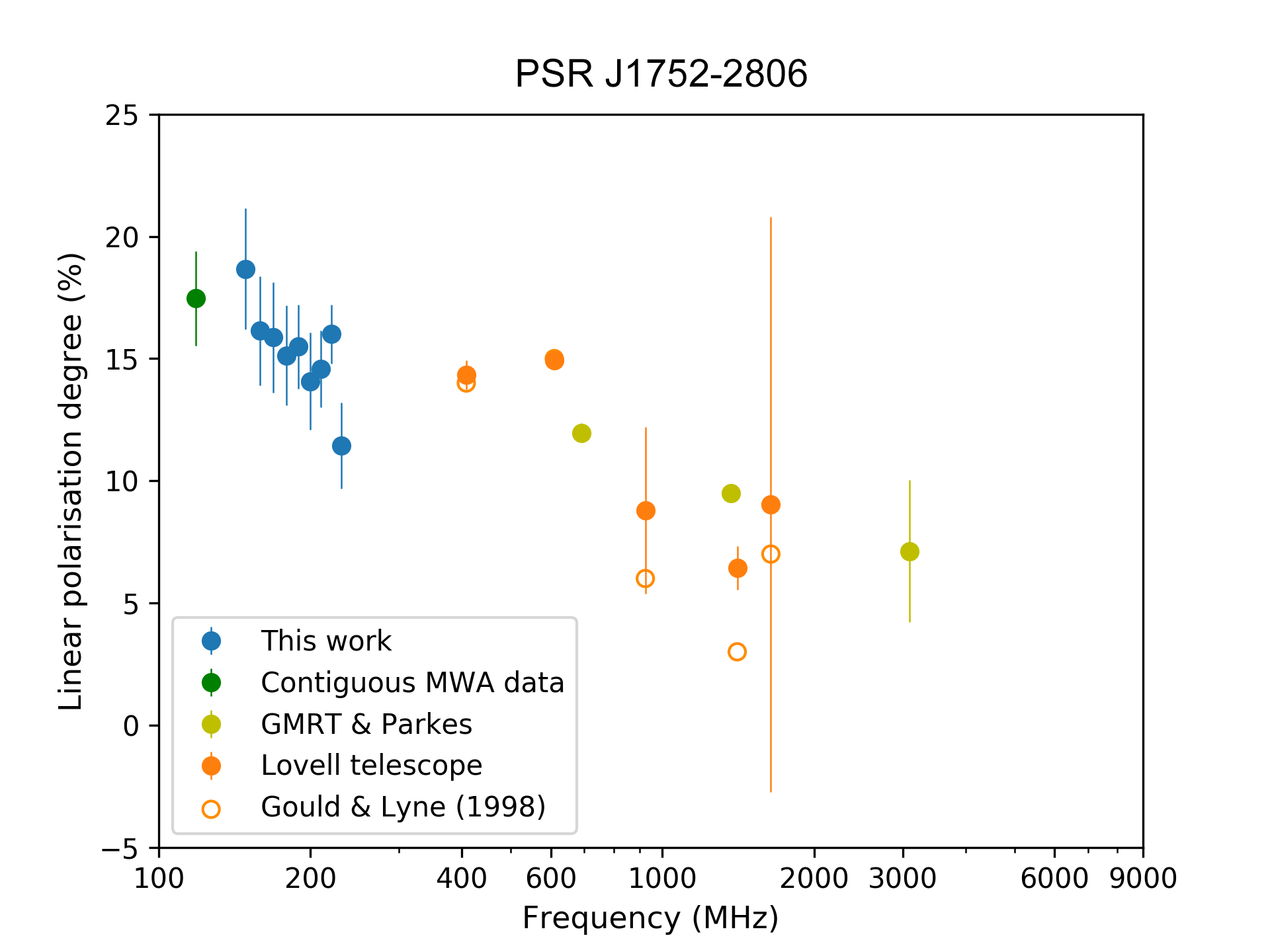}
\caption{Degree of linear polarisation as function of frequency for PSR J0742--2822 (left) and PSR J1752$-$2806 (right). Blue points indicate the degree of linear polarisation using the MWA data between 148 and 230\,MHz. Lime-green dots are from GMRT and Parkes data. Unfilled orange dots show the values published in \cite{Gould1998} and filled orange dots are calculated using the profiles in \cite{Gould1998}. For PSR J1752--2806, the dark-green dot shows the linear polarisation observed with the MWA 118\,MHz. For PSR J0742--2822 , the red line shows the depolarisation factor $\text{exp}(-2\lambda^{4}\delta\text{RM}^2)$ fit where $\delta$RM=0.13 rad\,m$^2$ (see Section \ref{sec:depol}).}\label{Fig:LoI}
\end{center}
\end{figure*}

Figure \ref{Fig:LoI} shows the measured degree of linear polarisation as a function of the observing frequency for PSR J0742--2822 (left) and PSR J1752$-$2806 (right). For PSR J1752$-$2806, the observed degree of linear polarisation increases towards lower frequencies, which is in line with the generally expected trend for pulsar emission. However, for PSR J0742--2822, the measured degree of linear polarisation peaks near 732\,MHz, whereas it decreases gradually at frequencies $\gtrsim$ 2\,GHz, and quite rapidly at frequencies $<$300\,MHz. This is in contradiction to the generally observed trend for many pulsars, where the degree of polarisation increases towards lower frequencies \citep[e.g. ][]{Manchester1971,Xilouris1996}. We note, therefore, that the frequency-averaged polarisation profile in Figure \ref{Fig:PA1} is purely for illustrative purpose, and that for any scientific interpretation, the profile evolution should be taken into account. For the frequency resolution that we have employed (10\,kHz), bandwidth depolarisation is expected only for $\rm |RM| \gtrsim 400~rad\,m^{-2}$. Moreover, as can be seen in Figure \ref{Fig:m1}, PSR J0742$-$2822 is also significantly scattered by the ISM. Therefore, it is possible that the depolarisation at low frequencies may due to scattering caused by the ISM.

There have been several studies in the literature discussing the impact of multi-path propagation in the ISM on the measured polarisation of pulsars \citep{Komesaroff1972,Li&Han2003,Karastergiou2009,Noutsos2009,Noutsos:2015}, which we will briefly review here. In early polarimetric studies of the Vela pulsar (PSR J0835$-$4510) from 300\,MHz to 1.4\,GHz, \cite{Komesaroff1972} measured a decrease in the degree of linear polarisation (at the pulse peak) and a smoother PA curve at their lower frequencies. While the smoother PA curve was interpreted as being due to scattering, the reduced degree in linear polarisation remained unexplained. Later, the work of \cite{Li&Han2003} investigated the impact of scattering on the PA curves, and suggested that the scattering can potentially flatten the PA curves. \cite{Karastergiou2009} further investigated this effect by comparing the observed profiles for pulsars whose PA curves adhere to the RVM with simulated scattered profiles. Their work suggested that very modest amounts of scattering, together with orthogonal jumps, can lead to significant distortion in the PA curve. \cite{Noutsos2009} investigated phase-resolved RM variations for a sample of pulsars using 1.4\,GHz Parkes data and suggested such variations can arise from interstellar scattering. This was further reinforced by \cite{Noutsos:2015} using LOFAR data, who also found that three of their pulsars (two of which exhibit significant scattering) show a decrease in the degree of linear polarisation at low frequencies ($<$200\,MHz).

For our MWA observations, the effective time resolution is mainly determined by the scattering time scale $\tau_{\rm s}$. For example, at a frequency of 180\,MHz, $\tau_{\rm s}$ is $\sim$\,6$\pm$3\,ms \citep{Kirsten2019}, which corresponds to 5 pulse phase bins given our sampling time resolution (see Table 2). This can potentially cause substantial smearing in polarisation profiles, leading to depolarisation. In such a scenario, a significant flattening of the PA curve is expected at lower frequencies, which is only seen in the trailing edge (in the scattering tail) of the pulse profile of PSR J0742$-$2822 (Figure \ref{Fig:PA1}). At earlier pulse phases, the PA curve shows structure in both the MWA and GMRT low-frequency data ($<$350\,MHz).

We considered the possibility that this depolarisation could be caused by short-term variability in the ionospheric RM. However, if that is the case, the depolarisation factor should relate to the total observation time that we used to generate the polarimetric profiles. To this end, we examined the depolarisation for both short (a few minutes) and long time scales (a few hours), given the constraints of our observations. Specifically, we compared the depolarisation between a) the addition of all five 5-minute short observations and each individual 5-minute observation; b) the first and second halves of each 5-minute observation. We found the depolarisation trends are consistent within the uncertainties. This indicates that the observed depolarisation is unlikely to be caused by the ionosphere. This is further reinforced by the fact that we see no significant depolarisation for PSR J1752$-$2806, and the ionosphere seems to show consistent behaviour between each pulsar's observing epochs.

Depolarisation can also arise due to propagation through turbulent plasma components that are irregularly magnetised \citep[e.g. ][]{Burn1966,Sokoloff1998}. In this case, stochastic Faraday rotation will depolarise the pulsar emission by a factor of $\text{exp}(-2\lambda^{4}\delta\text{RM}^2)$ \citep[e.g. ][]{Macquart2000a}, where $\lambda$ is the observing wavelength and $\delta\text{RM}$ is the fluctuation in the RM. By fitting the observed degree of linear polarisation as a function of frequency, we estimate the required RM fluctuation $\delta$RM = 0.13$\pm$0.02\,rad\,m$^{-2}$  (Figure \ref{Fig:LoI}). As previously mentioned, for PSR J0742$-$2822, $\tau_{\rm s} \sim 6 \pm 3$\,ms at 180\,MHz \citep{Kirsten2019}. \cite{Johnston1998} suggest that the scattering screen lies at a distance of $\sim$450 pc, associated with the Gum Nebula. In this case, the estimated size of the scattering disk is $\sim$ 33 AU.
Given that there is typically a greater amount of turbulent structures seen in the general direction of the Gum Nebula, it is possible that the observed RM fluctuation is caused by small-scale, random magneto-ionic components (associated with turbulence) within the scattering screen. Further investigations are required to examine this in more detail, and will be explored in a future publication.

\subsection{Limitations of data reduction and future work}
\label{sec:Discussleakage}

\begin{figure}
\begin{center}
\includegraphics[scale=0.5, angle=0]{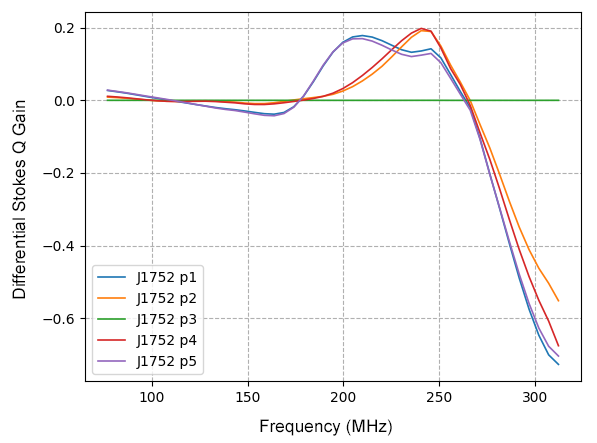}
\caption{An example of the deviation in normalised Stokes $Q$ gain value between the analytical and FEE beam models, as a function of observing frequency, for the directions of the five pointings towards PSR J1752--2806. }\label{Fig:beam_diff}
\end{center}
\end{figure}

For the work presented here, we have used the MWA's analytical beam model to calibrate and beamform the data. The analytical beam model is indeed a reasonably good approximation to the MWA's tile beam, if we consider total intensity alone.
For polarisation, the analytical beam model is acceptable (in terms of performance and stability) for observations at zenith angles \lapp 45$^{\circ}$ and at frequencies \lapp 270\,MHz.
Further developments have been made to the MWA tile beam model, including the recent fully-embedded element (FEE) primary beam model \citep{Sokolowski2017}. This is a significant improvement to the analytical model, and provides a more accurate prediction of the polarimetric response at large zenith angles. Therefore, using the FEE model in the calibration and beamforming will likely impart less instrumental polarimetric leakage to the data. This is illustrated in Figure \ref{Fig:beam_diff}, which shows the difference between the analytical and FEE beam models for the Stokes Q parameter, for example.
The increasing deviation above 270\,MHz indicates that the analytical beam model should be used with caution in this range.

In future, we plan to extend this work through improved polarimetric calibration by integrating the FEE beam model into our calibration and beamforming pipeline.
This is an important exercise, however, beyond the scope of the present work, given the much larger computational requirements of the FEE beam model.
We also plan to compare the MWA beamformed data with the imaging products \citep[e.g.][]{Lenc2017}, to provide further insights into the calibration and consistency between the observing modes, and to investigate any further corrections that could be applied to improve the polarimetric response.

\section{SUMMARY}
In this work, we have presented polarimetric studies of two bright southern pulsars, PSRs J0742--2822 and J1752--2806, using the MWA's recently-developed, full-polarisation, tied-array beam mode with high-time and -frequency resolution.
We have tested the response and stability of our calibration scheme over a range of zenith angles.
Our analysis shows that the polarimetric response of the MWA is reliable in the domain where the analytical beam model is a good approximation to reality. This is satisfied when the observing frequency is less than 270\,MHz and the zenith angle is less than 45\degree. As expected, observations closest to the zenith provide the highest signal-to-noise ratio pulse profiles.

For both PSRs J0742--2822 and J1752--2806, the profiles presented here are the first polarimetric profiles at frequencies below 300\,MHz. We have studied the polarimetric pulse profile evolution over the frequency range 97--230\,MHz using the MWA's large (non-contiguous) fractional bandwidth, and compared our results with the published work from higher frequency observations (up to $\sim$8.4\,GHz). For PSR J0742--2822, the measured degree of linear polarisation shows a rapid decrease at low frequencies, which is in contrast with the generally expected trend for pulsar emission. This depolarisation trend may be attributed to stochastic Faraday rotation across the scattering disk.

In the near future, we plan to extend these studies for a larger sample of pulsars, building on the earlier work of \cite{Xue2017}, in which we presented total intensity profiles of a large sample of pulsars using the MWA.
The polarimetric profiles of a large sample of pulsars using the MWA will also serve as a useful reference for calibrating SKA-Low pulsar data.

\begin{acknowledgements}
We thank Simon Johnston for kindly providing access to some of the unpublished data (for PSR J0742--2822), Jean-Pierre Macquart for the useful discussion on the depolarisation by the ISM, Willem van Straten for discussions relating to polarimetric analysis, and George Heald for providing the RM CLEAN code.
This scientific work makes use of the Murchison Radio-astronomy Observatory, operated by CSIRO. We acknowledge the Wajarri Yamatji people as the traditional owners of the Observatory site. Support for the operation of the MWA is provided by the Australian Government (NCRIS), under a contract to Curtin University administered by Astronomy Australia Limited. We acknowledge the Pawsey Supercomputing Centre which is supported by the Western Australian and Australian Governments. Parts of this research was supported by the Australian Research Council Centre of Excellence for All-sky Astrophysics (CAASTRO), through project number CE110001020. M. Xue is funded by the China Scholarship Council through the SKA PhD Scholarship project. BWM and SJM acknowledge the contribution of an Australian Government Research Training Program Scholarship in supporting this research.
\end{acknowledgements}

\bibliographystyle{pasa-mnras}
\bibliography{pulsar}

\end{document}